\documentclass[11pt,oneside,letterpaper]{article}

\usepackage{graphicx}
\DeclareGraphicsExtensions{.pdf,.png,.jpg}

\usepackage{color} 
\usepackage{parskip}

\usepackage{geometry}
\usepackage{caption}
\DeclareCaptionLabelFormat{number}{#2}
\usepackage{float}

\makeatletter
\renewcommand{\@biblabel}[1]{\quad#1.}
\makeatother

\usepackage{authblk}

\title{\vspace{1.5cm} \LARGE \bf Canalization of the evolutionary trajectory of the human influenza virus}

\author[1,2,$\dagger$]{Trevor Bedford}
\author[3,4]{Andrew Rambaut}
\author[1,2]{Mercedes Pascual}
\affil[1]{Department of Ecology and Evolutionary Biology, University of Michigan, Ann Arbor, MI, USA.} 
\affil[2]{Howard Hughes Medical Institute, University of Michigan, Ann Arbor, MI, USA.}
\affil[3]{Institute of Evolutionary Biology, University of Edinburgh, Edinburgh, UK.}
\affil[4]{Fogarty International Center, National Institutes of Health, Bethesda, MD, USA.}
\affil[$\dagger$]{Present address: Institute of Evolutionary Biology, University of Edinburgh, Edinburgh, UK.}
\date{}

\begin{document}

\renewcommand{\thefigure}{Figure~\arabic{figure}}
\renewcommand{\thetable}{Table~\arabic{table}}

\maketitle


\begin{abstract}
\begin{center}
\bf \normalsize Abstract
\end{center}

\vspace{0.5cm}

\noindent \normalsize Since its emergence in 1968, influenza A (H3N2) has evolved extensively in genotype and antigenic phenotype.  Antigenic evolution occurs in the context of a two-dimensional `antigenic map', while genetic evolution shows a characteristic ladder-like genealogical tree.  Here, we use a large-scale individual-based model to show that evolution in a Euclidean antigenic space provides a remarkable correspondence between model behavior and the epidemiological, antigenic, genealogical and geographic patterns observed in influenza virus.  We find that evolution away from existing human immunity results in rapid population turnover in the influenza virus and that this population turnover occurs primarily along a single antigenic axis.  Thus, selective dynamics induce a canalized evolutionary trajectory, in which the evolutionary fate of the influenza population is surprisingly repeatable and hence, in theory, predictable.
\end{abstract}

\vspace{0.5cm}
\pagebreak

\section*{Author Summary}

Each year, hundreds of millions become sick with the influenza virus and hundreds of thousands die from the infection.  After recovery from the flu, a person gains permanent immunity specific to the infecting strain.  However, owing to its RNA makeup, mutations occur rapidly to the virus genome.  Some of these mutations change the shape of proteins visible to the human immune system, and thus alter the virus's antigenic phenotype.  These mutations allow the virus to re-infect those who have previously recovered from an earlier strain, and thus will quickly spread through the virus population.  It is because of influenza's rapid antigenic evolution that the flu vaccine needs frequent updating.  However, despite strong pressure to evolve away from human immunity and to diversify in antigenic phenotype, influenza, and especially influenza A (H3N2), shows paradoxically limited genetic and antigenic diversity present at any one time.  Here, we propose a simple model of influenza that displays rapid evolution but low standing diversity and simultaneously accounts for the epidemiological, genetic, antigenic and geographical patterns displayed by the virus.  We find that in evolving directly away from existing human immunity, the virus severely limits its future evolutionary potential.

\section*{Introduction}

Epidemic influenza is responsible for between 250,000 and 500,000 global deaths annually, with influenza A (and in particular, A/H3N2) having caused the bulk of human mortality and morbidity \cite{flufactsheet}.  Influenza A (H3N2) has continually circulated within the human population since its introduction in 1968, exhibiting recurrent seasonal epidemics in temperate regions and less periodic transmission in the tropics.  During this time, H3N2 influenza has continually evolved both genetically and antigenically.  Most antigenic drift is thought to be driven by changes to epitopes in the hemagglutinin (HA) protein \cite{Nelson07NatRevGenet}.  Phylogenetic analysis of the genetic relationships among HA sequences has revealed a distinctive genealogical tree showing a single predominant trunk lineage and side branches that persist for only 1--5 years before going extinct \cite{Fitch97}.  This tree shape is indicative of serial replacement of strains over time; H3N2 influenza shows rapid evolution, but low standing genetic diversity.

This observation has remained puzzling from an epidemiological standpoint.  Antigenic evolution occurs rapidly and strong diversifying selection exists to escape from human immunity; why then do we see serial replacement of strains rather than continual genetic and antigenic diversification?  Indeed, simple epidemiological models show explosive diversity of genotype and phenotype over time \cite{Ferguson03,Tria05}.  Previous work has sought model-based explanations of the limited diversity of influenza, relying on short-lived strain-transcending immunity \cite{Ferguson03,Tria05}, complex genotype-to-phenotype maps \cite{Koelle06} or a limited repertoire of antigenic phenotypes \cite{Recker07}. 

Experimental characterization of antigenic phenotype is possible through the hemagglutination inhibition (HI) assay, which measures the cross-reactivity of HA from one virus strain to serum raised against another strain \cite{Hirst43}.  The results of many HI assays can be combined to yield a two-dimensional map, representing antigenic similarity and distance between strains as an easily visualized and quantified measure \cite{Smith04}.  The path traced across this map by influenza A (H3N2) from 1968 until present is largely linear, showing serial replacement of one strain by another; there are no major bifurcations of antigenic phenotype \cite{Smith04}.

Herein, we seek to simultaneously model the genetic and antigenic evolution of the influenza virus.  We represent antigenic phenotypes as points in a $N$-dimensional Euclidean space.  Based on the finding that a two-dimensional map adequately explains observed antigenic distance between strains \cite{Smith04}, we begin with antigenic phenotypes as points on a plane, but relax this assumption later on in the analysis.  After exposure to a virus, a host's risk of infection is proportional to the Euclidean distance between the infecting phenotype and the closest phenotype in the host's immune history.  Mutations perturb antigenic phenotype, moving phenotype in a random radial direction and for a randomly distributed distance.  We implemented this geometrical model in a large-scale individual-based simulation intended to directly model the antigenic map and genealogical tree of the global influenza population.  The simulation includes multiple host populations with different seasonal forcing, hosts with complete immune histories of infection, and viruses with antigenic phenotypes.  As the simulation proceeds, infections are tracked and a complete genealogy connecting virus samples is constructed.  Results shown here are for a single representative simulation of 40 years of virus evolution in a population of 90 million hosts.

\section*{Results}

\begin{figure}[H]
	\centering
	\makebox[\textwidth]{	
		\includegraphics{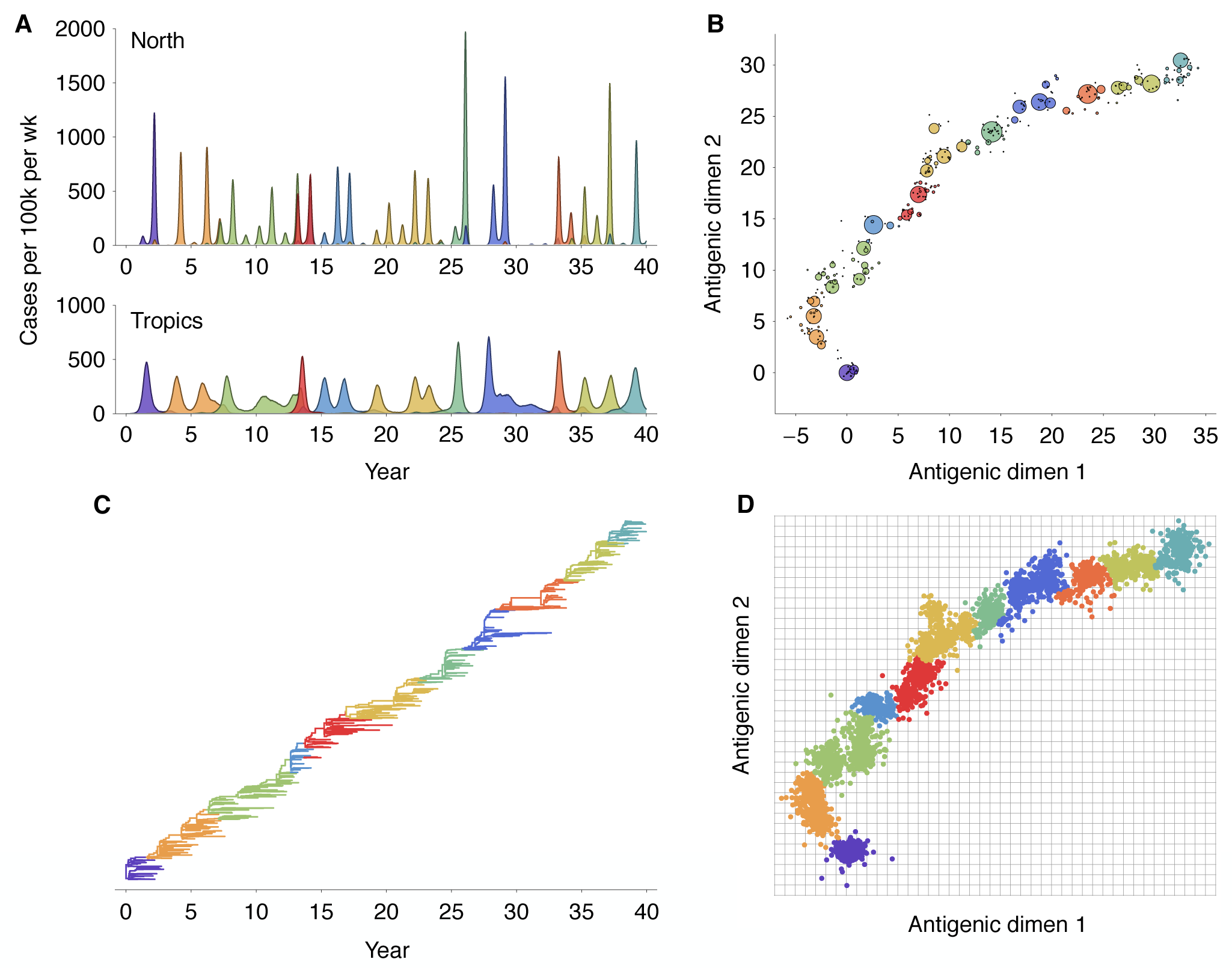}
	}
	\caption{\textbf{Simulation results showing epidemiological, antigenic and genealogical dynamics}. (A) Weekly timeseries of incidence of viral infection in north and tropics regions.  (B) Two-dimensional antigenic phenotypes of 5943 viruses sampled over the course of the simulation. Each discrete virus phenotype is shown as a bubble, with bubble area proportional to the number of times this phenotype was sampled.  (C) Genealogical tree depicting the infection history of 376 samples from the virus population.  Parent/offspring relationships were tracked over the course of the simulation, giving a direct observation of the genealogy rather than a phylogenetic inference.  (D) Antigenic map depicting phenotypes of 5943 viruses sampled over the course of the simulation.  To construct the map, noise was added to each sample and the resulting observations grouped into 11 clusters and colored accordingly.  Grid lines show single units of antigenic distance.  Cluster assignments were used to color all panels in a consistent fashion.}
	\label{evol}
\end{figure}

The virus persists over the course of the 40-year simulation, infecting a significant fraction of the host population through annual winter epidemics in temperate regions and through less periodic epidemics in the tropics (\ref{evol}A).  Across replicate simulations, we observe average yearly attack rates of 6.8\% in temperate regions and rates of 7.1\% in the tropics, comparable with estimated attack rates of influenza A (H3N2) of 3--8\% per year \cite{Monto93,Koelle09}.  Over the course of the simulation, the virus population evolves in antigenic phenotype exhibiting, at any point, a handful of highly abundant phenotypes sampled repeatedly and a large number of phenotypes appearing at low abundance (\ref{evol}B).  By including measurement noise on antigenic locations, we approximate an experimental antigenic map of H3N2 influenza (\ref{evol}D).  The appearance of clusters in the antigenic map comes from the regular spacing of high abundance phenotypes combined with measurement noise.  Over time, clusters of antigenically similar strains are replaced by novel clusters of more advanced strains (\ref{phenotypes}A).  Across replicate simulations, clusters persist for an average of 5.0 years measured as the time it takes for a new cluster to reach 10\% frequency, peak and decline to 10\% frequency.  The transition between clusters occurs quickly, taking an average of 1.8 years.

Remarkably, although antigenic phenotype is free to mutate in any direction in the two-dimensional space, selection pressures force the virus population to move in nearly a straight line in antigenic space (\ref{evol}B).  Across replicate simulations, 94\% of the variance of antigenic phenotype can be explained by a single dimension of variation.  This mirrors the empirical results showing a largely linear antigenic map for H3N2 influenza isolates from 1968 to 2003 \cite{Smith04}.  Because of the primarily one-dimensional movement, antigenic distance from the original phenotype increases nearly linearly with time (\ref{phenotypes}B).  Antigenic evolution occurs in a punctuated fashion; periods of relative stasis are interspersed with more rapid antigenic change (\ref{phenotypes}B).  Antigenic and epidemiological dynamics show a fundamental linkage, so that large jumps of antigenic phenotype result in increased rates of infection (\ref{evol}, \ref{driftvsinc}).  In general, evolution via many smaller mutations results in a smoother antigenic map and less variation in yearly epidemics (\ref{incmaptree_smooth}), while evolution via rare mutations of large effect exhibits a more clustered antigenic map and wider variation in seasonal incidence (\ref{incmaptree_rough}).

The genealogical tree connecting the evolving virus population appears characteristically sparse with pronounced trunk and short-lived side branches  (\ref{evol}C).  This tree shape is reflected in low levels of standing diversity; across replicates, an average of 5.68 years of evolution separate two randomly sampled viruses in the population.  This level of diversity matches what is observed in phylogenies of influenza A (H3N2) \cite{Rambaut08}.  A spindly genealogical tree is indicative of population turnover, wherein novel antigenic phenotypes continually replace more primitive `spent' phenotypes, purging their genealogical diversity.  In general, natural selection reduces effective population size and genealogical diversity \cite{BedfordBMC11}.  Here, by comparing mutations occurring on trunk branches vs.\ mutations occurring on side branches, we find evidence for pervasive positive selection for antigenic change (\ref{mktable}).  Trunk mutations tend to push antigenic phenotype forward along the line of primary antigenic variation (\ref{mutspectrum}).  We find a roughly linear relationship between the antigenic effect of a mutation and the likelihood that this mutation becomes incorporated into the trunk (\ref{probtrunk}).  Additionally, we find that trunk mutations occur at strikingly regular intervals, with less variation of waiting times than expected under a simple random process (\ref{waittimes}).  There is a relative scarcity of mutation events occurring in intervals under 1 year and a relative excess of a mutation events occurring in 2--3 year intervals (\ref{waittimes}).

\begin{figure}[H]
	\centering
	\includegraphics{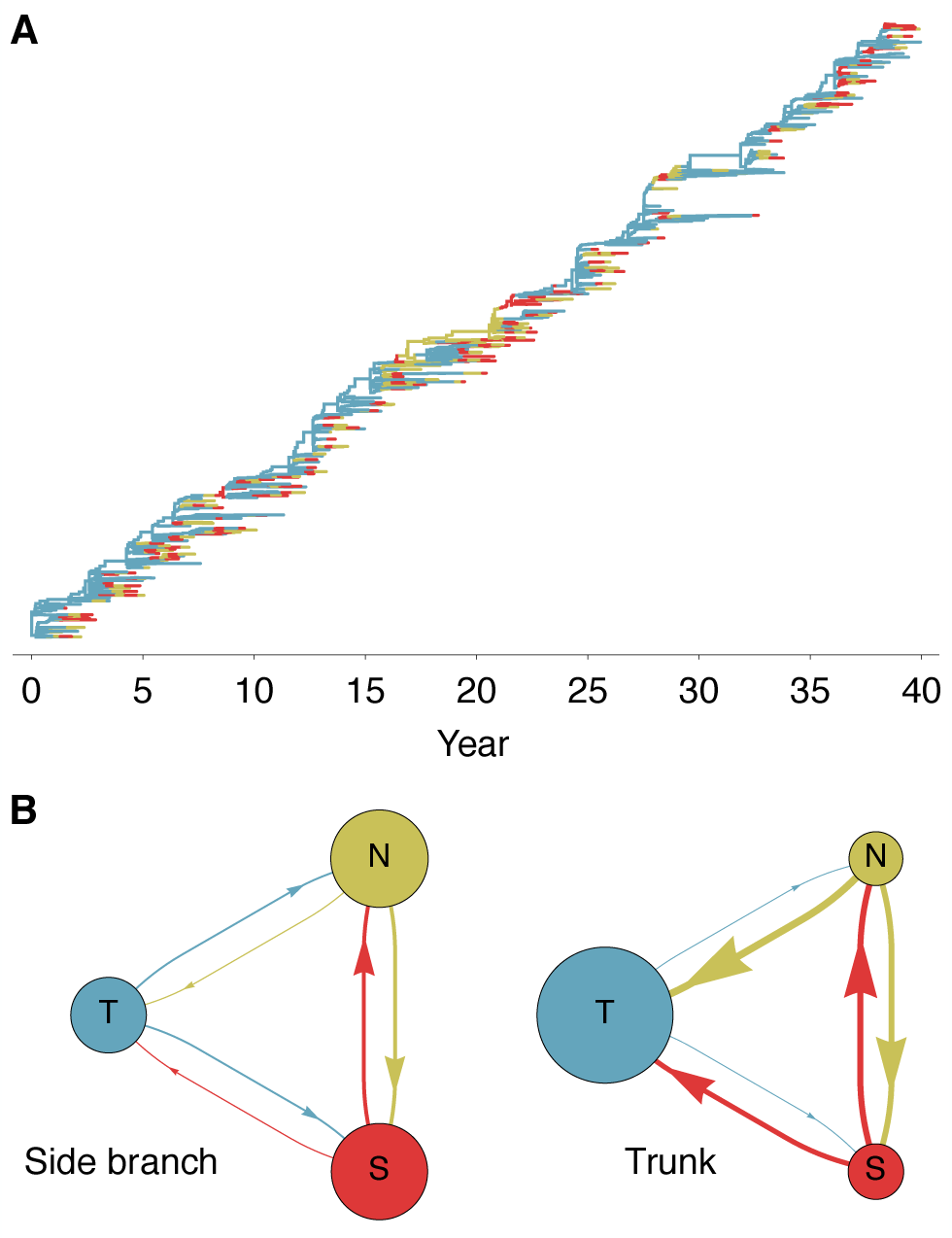}
	\caption{\textbf{Patterns of geographic movement of virus lineages}. (A) Evolutionary relationships among 376 viruses sampled evenly through time colored by geographic location. Lineages residing in the north (N), south (S) and tropics (T) are colored yellow, red and blue respectively. (B) Observed migration rates between regions on side branch lineages (left) and on trunk lineages (right). Arrows denote movement of lineages and arrow width is proportional to migration rate. Circle area is proportional to the expected stationary frequency of a region given the observed migration rates.  In both cases, migration rates are calculated across 80 replicate simulations.}
	\label{spatial}
\end{figure}

The genealogical tree also contains detailed information on the history of migration between regions.  We find that, consistent with empirical estimates \cite{Russell08,Bedford10}, the trunk resides primarily within the tropics, where seasonal dynamics are less prevalent (\ref{spatial}A).  Across replicate simulations, we observe 72\% of the trunk's history within the tropics and 28\% within temperate regions.  With symmetric host contact rates and equal host population sizes, and without seasonal forcing, we would expect trunk proportions of one third for each region.  We calculated rates of migration based on observed event counts across replicate simulations, separating region-specific rates on side branches from region-specific rates on trunk branches.  We find that migration patterns on side branches are close to symmetric, with similar rates between all regions, while migration patterns on trunk branches are highly asymmetric, with high rates of movement between temperate regions and from temperate regions into the tropics (\ref{spatial}B).  Extrapolating from these rates, we arrive at an expected stationary distribution of 76\% tropics and 24\% temperate regions, in line with the observed residency patterns of the trunk.  It may at first seem counter-intuitive to see higher rates of movement from the temperate regions into the tropics along trunk branches, but it makes sense when thought of in terms of conditional probability.  Only those lineages that migrate into the tropics or those lineages which rapidly migrate between the north and south have a chance at becoming the trunk lineage, while lineages that remain within the temperate regions are doomed to extinction.  

These findings suggest that persistence and migration are fundamentally connected and have important implications for future phylogeographic analyses.  Russell et al.\ \cite{Russell08} emphasize a source-sink model of movement of the HA protein of influenza A (H3N2) based on their finding of a trunk lineage residing within China and the Southeast Asian tropics.  Whereas, Bedford et al.\ \cite{Bedford10} emphasize a global metapopulation model based on phylogenetic inference of migration rates across the entire tree.  Our results suggest that both scenarios are simultaneously possible; side branches may be highly volatile moving rapidly and symmetrically between regions, while the trunk lineage may be more stable remaining within a region (or within a highly connected network of regions) that has more persistent transmission.  In light of these results, we suggest that future work on the phylogeography of influenza take into account trunk vs.\ side branch differences in migration patterns.

\section*{Discussion}

\subsection*{Correspondence between model and data}

Although multiple epidemiological/evolutionary mechanisms have been proposed to explain the restricted genetic diversity and rapid population turnover of influenza A (H3N2) \cite{Ferguson03,Tria05,Koelle06,Recker07}, our results show that a simple model coupling antigenic and genealogical evolution exhibits broad explanatory power.  We find a strong correspondence between the antigenic and genealogical patterns generated by our model (\ref{evol}) and patterns of genetic and antigenic evolution exhibited by influenza A (H3N2) \cite{Fitch97,Smith04}.  Our model suggests that punctuated antigenic evolution need only be explained by a lack of mutational opportunity and predicts that more detailed classification of influenza strains will support a relatively small number of predominant phenotypes (\ref{evol}B).  We suggest that a large proportion of intra-cluster variation in the observed antigenic map is due to experimental noise, rather than each strain possessing a unique antigenic location.  Additionally, our model accurately predicts the contrasting dynamics of other types/subtypes of influenza.  We find that lowering mutation size/effect or lowering intrinsic $R_0$ results in decreased incidence, slower antigenic movement and greater genealogical diversity, all distinguishing characteristics of H1N1 influenza and influenza B (\ref{h1n1_mut}, \ref{h1n1_r0}).  

In our model, when antigenic phenotype remains static, there may be multiple consecutive seasons without appreciable incidence (\ref{evol}A), a pattern apparently absent from H3N2 influenza \cite{Finkelman07}.  We suggest that any model exhibiting punctuated evolution broadly consistent with the punctuated change seen in the experimental antigenic map will show similar patterns of incidence.  We can `fix' the incidence patterns, but at the cost of too smooth an antigenic map (\ref{incmaptree_smooth}).  Evolutionary patterns of the neuraminidase (NA) protein may provide an explanation.  Epitopes in the HA and NA proteins are jointly responsible for determining antigenicity \cite{Nelson07NatRevGenet}, and it is now clear that levels of adaptive evolution are similar between HA and NA \cite{Bhatt11}.  Thus, changes in NA may be driving incidence patterns as well, resulting in an observed timeseries of incidence partially divorced from the antigenic map of HA.

It remains a central question as to the extent that short-lived strain-transcending immunity is responsible for influenza's limited diversity and spindly genealogical tree \cite{Ferguson03,Tria05}.  Our findings suggest a possible resolution.  Although lacking short-lived immunity, our model shows a detailed correspondence to both the antigenic map and genealogical tree of H3N2 influenza.  If an antigenic map were to show a deep bifurcation, where two viral lineages move in different antigenic directions, then we would expect the same bifurcation to be evident in the genealogical tree.  Short-lived strain-transcending immunity provides a mechanism by which lineages may diverge in antigenic phenotype, but still show epidemiological interference.  This mechanism would explain a situation where bifurcations emerge in the antigenic map, but competition results in the extinction of divergent antigenic lineages.  The empirical antigenic map \cite{Smith04} suggests that this is not the case; one cluster leads to another cluster in orderly succession and there is never competition between antigenically distant clusters.  This supports the hypothesis that antigenic evolution is primarily limited by a lack of mutational availability.  This is not to say that short-lived strain-transcending immunity is not present; observed interference between subtypes \cite{Ferguson03,Goldstein11} and evolution at CTL epitopes \cite{Voeten00} provides substantial evidence for its existence.  Instead, we suggest that short-lived strain-transcending immunity does not automatically generate antigenic maps and genealogical trees consistent with empirical evidence.

\subsection*{Linear antigenic movement}

\begin{figure}[H]
	\centering
	\includegraphics{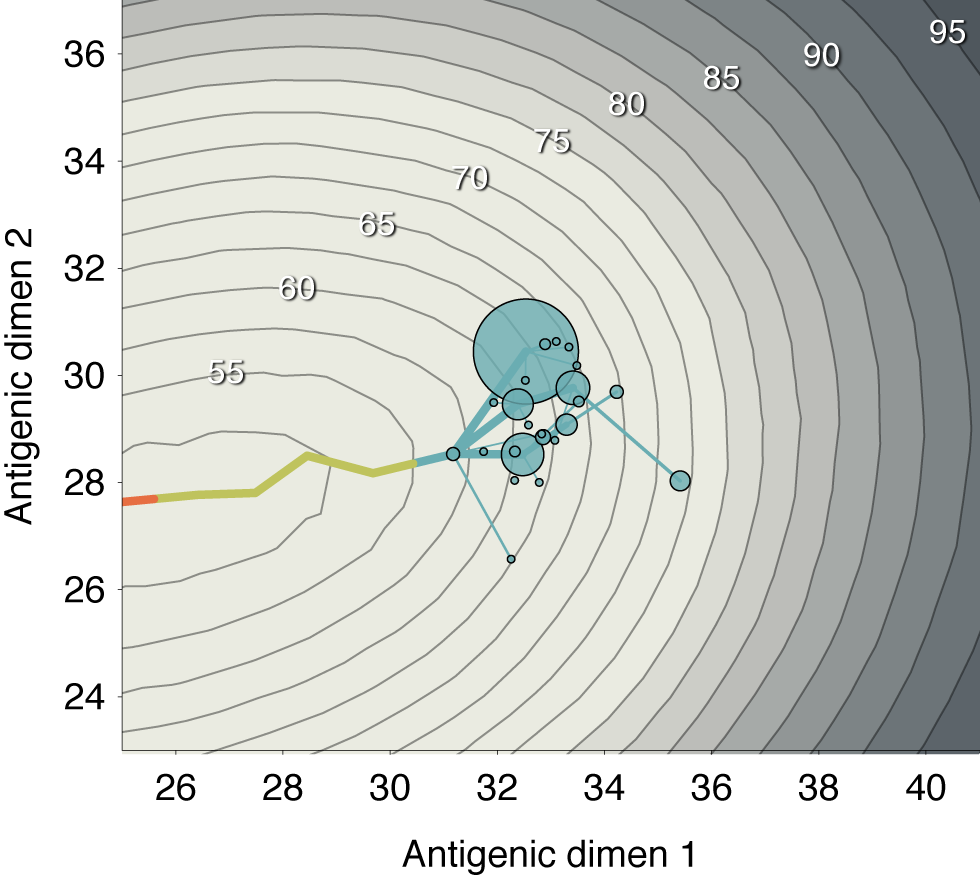}
	\caption{\textbf{Host immunity and antigenic history of the virus population}.  Contour lines represent the state of host immunity at the end of the 40-year simulation.  They show the mean risk of infection (as a percentage) after a random host in the population encounters a virus bearing a particular antigenic phenotype.  Contour lines are spaced in intervals of 2.5\%. Bubbles represent a sample of antigenic phenotypes present at the end of the 40-year simulation.  The area of each bubble is proportional to the number of samples with this phenotype.  Lines leading into these bubbles show past antigenic history.  The current phenotypes rapidly coalesce to a trunk phenotype.  The movement of the virus population from the left to the center of the figure can be seen from the antigenic history.  At the end of the simulation several virus phenotypes exist with similar antigenic locations; all of these phenotypes lie significantly ahead of the peak of host immunity.}
	\label{immunity}
\end{figure}

It would seem possible for one viral lineage to move in one antigenic direction, while another lineage moves tangentially, eventually resulting in two non-interacting viral lineages.  Instead, we find that only movement in a single antigenic direction is favored.  The origins of this pattern can be seen in the interaction between virus evolution and host immunity (\ref{immunity}).  As the virus population evolves forward it leaves a wake of immunity in the host population, and evolution away from this immunity results in the canalization of the antigenic phenotype; mutations that continue along the line of primary antigenic variation will show a transmission advantage compared to more tangential mutations.  

Following the work of Smith et al.\ \cite{Smith04}, it remained an open question of why a two-dimensional map should explain the antigenic variation of H3N2 influenza.  Although the authors astutely speculated that ``there is a selective advantage for clusters that move away linearly from previous clusters as they most effectively escape existing population-level immunity, and this is a plausible explanation for the somewhat linear antigenic evolution in regions of the antigenic map.''  This hypothesis remained to be tested.  Here, we show from a simple model of epidemiology and evolution that a linear trajectory of antigenic evolution dynamically emerges due to basic selective pressures.  This result simultaneously explains the linear pattern of antigenic drift \cite{Smith04} and the characteristically spindly genealogical tree \cite{Fitch97} exhibited by influenza A (H3N2).

To consider to what extent these results were contingent on the dimensionality of the underlying antigenic model, we further implemented our model in a 10-dimensional antigenic space.  Here, mutations occur as 10-spheres, but the distance moved by a mutation is the same as in the previous two-dimensional formulation.  We arrive at nearly the same results with this model; principal components analysis shows that the first and second dimensions of variation account for 87\% and 7\%, respectively, of the total variance (\ref{10dgrid}).  Thus, our model predicts that future work probing mutational effects will support an underlying high-dimensional antigenic space, even though a two-dimensional map is sufficient to explain observed antigenic relationships among strains.

\subsection*{Winding back the tape}

\begin{figure}[H]
	\centering
	\makebox[\textwidth]{	
		\includegraphics{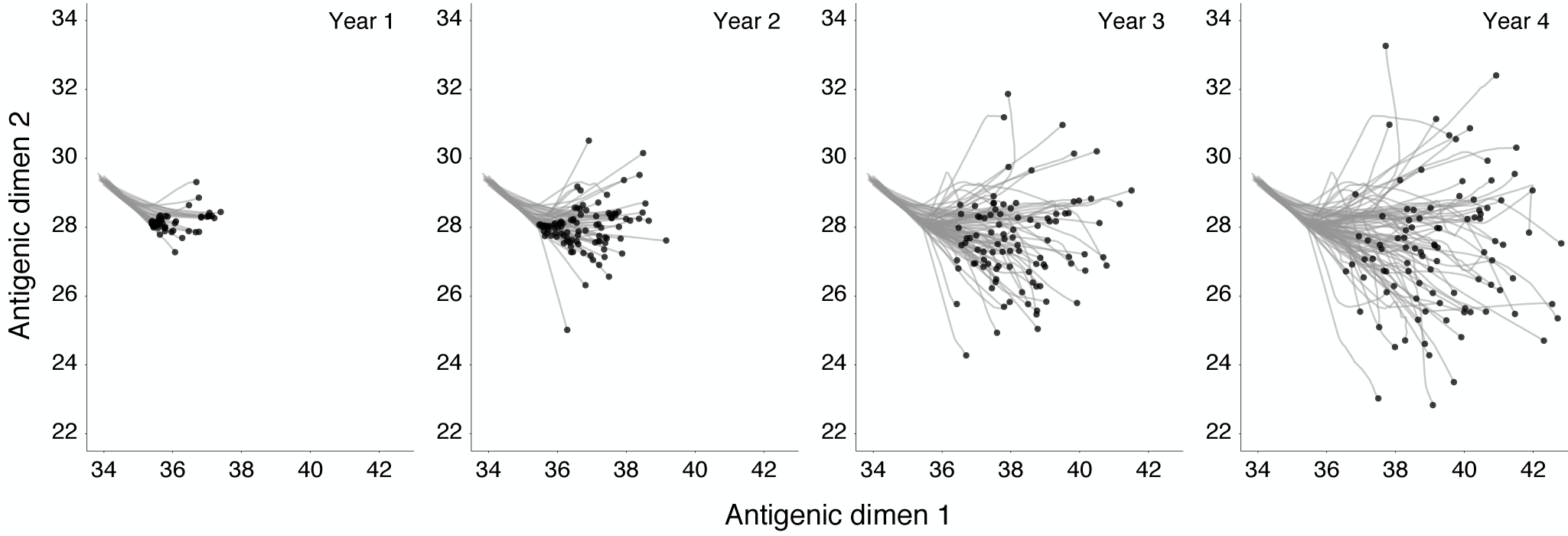}
	}
	\caption{\textbf{Antigenic phenotypes over the course of 4 years of evolution across 100 replicate simulations starting from identical initial conditions}.  Replicate simulations were initialized with the end state of the  40-year simulation shown in \ref{evol}.  Each panel shows an additional year of evolution, with black points representing the mean antigenic phenotypes of the 100 replicate simulations and gray lines representing the history of each mean antigenic phenotype.}
	\label{replicateevol}
\end{figure}

\begin{figure}[H]
	\centering
	\includegraphics{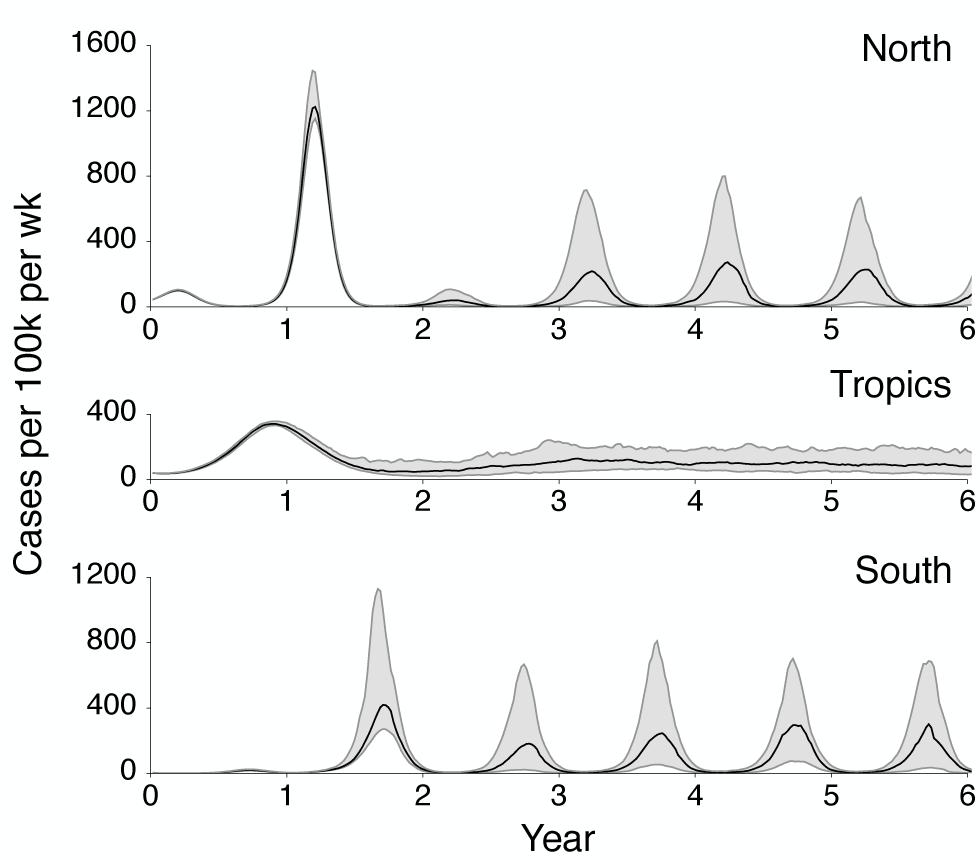}
	\caption{\textbf{Timeseries of incidence across 100 replicate simulations with identical initial conditions.} Panels show incidence in the North, Tropics and South regions over the course of 6 years.  Solid black lines represent the median weekly incidence across the 100 replicate simulations, while gray intervals represent the interquartile range across simulations.  There is little variability for the first year of replicate simulations.  Replicate simulations were initialized with the end state of the 40-year simulation shown in \ref{evol}.}
	\label{replicateinc}
\end{figure}

It seems clear that, in our model, selection reduces the degrees of freedom of antigenic evolution.  In light of this, we wanted to examine the degree of stochasticity in replicated evolutionary trajectories, and thereby test what happens when we ``wind back the tape'' \cite{GouldWonderfulLife} on the evolution of the virus.  We ran 100 replicate simulations, each starting from the endpoint of the original 40-year simulation (\ref{replicateevol}, \ref{replicateinc}).  Initially, we find a great detail of repeatability; during the first year of evolution, every replicate virus population undergoes a similar antigenic transition (\ref{replicateevol}), resulting in a repeatable peak in northern hemisphere incidence (\ref{replicateinc}).  After three years, repeatability has mostly disappeared, with antigenic phenotype and incidence appearing highly variable across replicates (\ref{replicateevol}, \ref{replicateinc}).  The 1--2 year timescale of repeatability can be explained by the presence of standing antigenic variation.  In the initial virus population, there are several novel antigenic variants present at low frequency (\ref{immunity}), one of which, without fail, comes to predominate the virus population.  

We see that the initial evolutionary trajectory, during which time standing variation plays out, is highly repeatable, and thus predictable given enough information and the right methods of analysis.  However, prediction of longer-term evolutionary scenarios will necessarily be difficult or impossible except in a vague sense.  Through careful surveillance efforts and genetic and antigenic characterization of influenza strains, the World Health Organization makes twice-yearly vaccine strain recommendations \cite{Barr10}.  It should be possible to combine these sorts of modeling approaches with surveillance data to gauge the likelihood that a sampled variant will spread through the population.

Recent work on empirical fitness landscapes has shown that natural selection follows few mutational paths \cite{Weinreich06}.  The spindly genealogical tree and the almost linear serial replacement of influenza strains has remained a puzzling phenomenon.  We suggest that the evolutionary and epidemiological dynamics displayed by the influenza virus may simply be explained as an outgrowth of selection to avoid host immunity.  Natural selection can only `see' one step ahead, and so sacrifices long-term gains for short-term advantages.  The result is a canalized evolutionary trajectory lacking antigenic diversification.

\section*{Materials and Methods}

\subsection*{Transmission model}

To characterize the joint epidemiological, genealogical, antigenic and spatial patterns of influenza, we implemented a large-scale individual-based model.  This model consists of daily time steps, in which the states of hosts and viruses are updated.  Hosts may be born, may die, may contact other hosts allowing viral transmission, or may recover from infection.  Viruses may mutate in antigenic phenotype.  Each simulation ran for 40 years of model time.  

Hosts in this model are divided between three regions: North, South and Tropics.  There are 30 million hosts within each of the three regions, giving $N = 9 \times 10^{7}$ hosts.  Host population size remains fixed at this number, but vital dynamics cause births and deaths of hosts at a rate of $1 / 30$ years $= 9.1 \times 10^{-5}$ per host per day.  Within each region, transmission proceeds through mass-action with contacts between hosts occurring at a rate of $\beta = 0.36$ per host per day.  Regional transmission rates in temperate regions vary according to sinusoidal seasonal forcing with amplitude $\epsilon = 0.15$ and opposite phase in the North and in the South.  Transmission rate does not vary over time in the Tropics.  Transmission between region $i$ and region $j$ occurs at rate $m\,\beta_i$, where $m=0.001$ and is the same between each pair of regions and $\beta_i$ is the within-region contact rate.   Hosts recover from infection at rate $\nu = 0.2$ per host per day, so that $R_0$ in a naive host population is 1.8.  There is no super-infection in the model.

Each virus possesses an antigenic phenotype, represented as a location in Euclidean space.  Here, we primarily use a two-dimensional antigenic location.  After recovery, a host `remembers' the phenotype of its infecting virus as part of its immune history.  When a contact event occurs and a virus attempts to infect a host, the Euclidean distance from infecting phenotype $\phi_v$ is calculated to each of the phenotypes in the host immune history $\phi_{h_1}, \dots, \phi_{h_n}$.  Here, one unit of antigenic distance is designed to correspond to a twofold dilution of antiserum in a hemagglutination inhibition (HI) assay \cite{Smith04}. The probability that infection occurs after exposure is proportional to the distance $d$ to the closest phenotype in the host immune history.  Risk of infection follows the form $\rho = \textrm{min}\{d\,s,1\}$, where $s=0.07$.  Cross-immunity $\sigma$ equals $1-\rho$.  The initial host population begins with enough immunity to slow down the initial virus upswing and place the dynamics closer to their equilibrium state; initial $R$ was 1.28.

Our model follows Gog and Grenfell \cite{Gog02} in representing antigenic distance as distance between points in a geometric space.  By forcing one-dimension to directly modulate $\beta$, Gog and Grenfell find an orderly replacement of strains.  Kryazhimskiy et al. \cite{Kryazhimskiy07} use a two-dimensional strain-space, but enforce a cross-immunity kernel that directly favors moving along a diagonal line away from previous strains.  Our model does not `build in' the one-dimensional direction of antigenic drift, which instead emerges dynamically from competition among strains.

The initial virus population consisted of 10 infections each with the identical antigenic phenotype of $\{0,0\}$.  Over time viruses evolve in antigenic phenotype.  Each day there is a chance $\mu = 10^{-4}$ that an infection mutates to a new phenotype.  This mutation rate represents a phenotypic rate, rather than genetic mutation rate, and can be thought of as arising from multiple genetic sources.  When a mutation occurs, the virus's phenotype is moved in a completely random direction $\sim \textrm{Uniform}(0,360)$ degrees. Mutation size is sampled from the distribution $\sim \textrm{Gamma}(\alpha,\beta)$, where $\alpha$ and $\beta$ are chosen to give a mean mutation size of 0.6 units and a standard deviation of 0.4 units.  This distribution is parameterized so that mutation usually has little effect on antigenic phenotype, but occasionally has a large effect.  This is similar to the neutral networks implemented by Koelle et al. \cite{Koelle06}, wherein most amino acid changes result in little decrease to cross-immunity between strains, but some changes result in large jumps in cross-immunity.

\subsection*{Model output}

Daily incidence and prevalence are recorded for each region.  During the course of the simulation, samples of current infections are taken from the evolving virus population at a rate proportional to prevalence.  Each viral infection is assigned a unique ID, and in addition, infections have their phenotypes, locations and dates of infection recorded.  In this model, viruses lack sequences and so standard phylogenetic inference of the evolutionary relationships among strains is impossible.  Instead, the viral genealogy is directly recorded.  This is made possible by tracking transmission events connecting infections during the simulation; infections record the ID of their `parent' infection.  Proceeding from a sample of infections, their genealogical history can be reconstructed by following consecutive links to parental infections.  During this procedure, lineages coalesce to the ancestral lineages shared by the sampled infections, eventually arriving at the initial infection introduced at the beginning of the simulation.  Commonly, phylodynamic simulations generate sequences that are subsequently analyzed with phylogenetic software to produce an estimated genealogy \cite{Ferguson03,Koelle06,Koelle10}.  This step of phylogenetic inference is imperfect and computationally intensive, and by side-stepping phylogenetic reconstruction we arrive at genealogies quickly and accurately.  Other authors have implemented similar tracking of infection trees \cite{Volz09,Odea11}.  This genealogy-centric approach makes many otherwise difficult calculations transparent, such as calculating lineage-specific region-specific migration rates (\ref{spatial}) and lineage-specific mutation effects (\ref{mutspectrum}).

Infections are sampled at a rate designed to give approximately 6000 samples over the course of the simulation, with genealogies constructed from a subsample of approximately 300 samples.  The results presented in \ref{evol} represent a single representative model output; one hundred replicate simulations were conducted to arrive at statistical estimates. 

\subsection*{Parameter selection and sensitivity analysis}

Estimating what the basic reproductive number $R_0$ for seasonal influenza would be in a naive population is notoriously difficult.  Season-to-season estimates of effective reproductive number $R$ for the USA and France gathered from mortality timeseries display an interquartile range of 0.9--1.8 \cite{Chowell08}.  Geographic spread within the USA suggests an average seasonal $R$ of 1.35 \cite{Viboud06}.  These estimates of $R$ will be lower than the $R_0$ of influenza due to the effects of human immunity.  We assumed $R_0$ of 1.8, consistent with the upper range of seasonal estimates.  Duration of infection was chosen based on patterns of viral shedding shown during challenge studies \cite{Carrat08}.  The linear form of the risk of infection and its increase as a function of antigenic distance $s$ was chosen as 0.07 based on experimental work on equine influenza \cite{Park09} and from studies of vaccine effectiveness \cite{Gupta06}.  Between-region contact rate $m$ was chosen to yield a trunk lineage that resides predominantly in the tropics.  With much higher rates of mixing, the trunk lineage ceases to show a preference the tropics, and with much lower rates of mixing, particular seasons in the north and the south will often be skipped.  The amplitude of seasonal forcing $\epsilon$ was chosen to be just large enough to get consistent fade-outs in the summer months and is consistent with empirical estimates \cite{Truscott11}.

Mutational parameters were based, in part, on model behavior.  We assumed 10 amino acid sites involved in antigenicity, each mutating at a rate of $10^{-5}$ \cite{Rambaut08} to give a phenotypic mutation rate $\mu = 10^{-4}$ per infection per day.  We chose mutational effect parameters ($\textrm{mean} = 0.6$, $\textrm{sd} = 0.4$) that would give suitably fast rates of antigenic evolution corresponding to approximately 1.2 units of antigenic change per year, while simultaneously giving clustered patterns of antigenic evolution  \cite{Smith04}.  Similar outcomes are possible under a variety of parameterizations.  If mutations are more common ($\mu = 3 \times 10^{-4}$) and show less variation in effect size ($\textrm{mean} = 0.6$, $\textrm{sd} = 0.2$), then antigenic drift occurs in a more continuous fashion, resulting in less variation in seasonal incidence and a smoother distribution of antigenic phenotypes (\ref{incmaptree_smooth}).  If mutations are less common ($\mu = 5 \times 10^{-5}$) and show more variance in effect ($\textrm{mean} = 0.7$, $\textrm{sd} = 0.5$), then antigenic change occurs in a more punctuated fashion (\ref{incmaptree_rough}).  Basic reproductive number $R_0$ can be traded off with mutational parameters to some extent.  Less mutational input and higher $R_0$ will give similar patterns of antigenic drift and seasonal incidence.  Similarly, Kucharski and Gog \cite{Kucharski11} find that increasing $R_0$ results in increased rates of emergence of antigenically novel strains.

In 20 out of the 100 replicate simulations, we observed a major bifurcation of antigenic phenotype and a consequent increase in incidence and genealogical diversity.  These simulations were removed from the analysis.   Similar to Koelle et al. \cite{Koelle11}, we assume that although the historical evolution of H3N2 influenza followed the path of a single lineage, it could have included a major bifurcation.  Further work in these directions will help to determine the likelihoods of single lineage vs.\ bifurcating scenarios.

\subsection*{Antigenic map}

Antigenic phenotypes are modeled as discrete entities on the Euclidean plane; multiple samples have the same antigenic location.  However, in the empirical antigenic map of influenza A (H3N2), each strain appears in a unique location \cite{Smith04}.  We would argue that some of this pattern comes from experimental noise.  Indeed, Smith et al. \cite{Smith04} find that observed measurements and measurements predicted from the map differ by an average of 0.83 antigenic units with a standard deviation of 0.67 antigenic units.  We take this as a proxy for experimental noise and add jitter to each sampled antigenic phenotype by moving it in a random direction for an exponentially distributed distance with mean of 0.53 antigenic units.  If two samples with the same underlying antigenic phenotype are jittered in this fashion, the distance between them averages 0.83 antigenic units with a standard deviation of 0.64 units.

We added noise to each of the 5943 sampled viruses in this fashion resulting in an approximated antigenic map (\ref{evol}D).  Virus samples were then clustering following standard clustering algorithms.  We tried clustering by the $k$-means algorithm and also agglomerative hierarchical clustering with a variety of linkage criterion.  We found that clustering by Ward's criterion consistently outperformed other methods, when measured in terms of within-cluster and between-cluster variances.  However, the exact clustering algorithm had little effect on our overall results.

\subsection*{Acknowledgments} 
We would like to thank Sarah Cobey, Aaron King, Pejman Rohani and the attendees of the 2011 RAPIDD Workshop on Phylodynamics for helpful discussion.  We would also like to thank Ed Baskerville and Daniel Zinder for programming advice.  The term `canalization' was originally suggested by Micaela Martinez-Bakker. 

\subsection*{Funding} 
TB is supported by the Howard Hughes Medical Institute and by the European Molecular Biology Organization.  AR works as a part of the Interdisciplinary Centre for Human and Avian Influenza Research (ICHAIR) and the University of Edinburgh's Centre for Immunity, Infection and Evolution (CIIE).  MP is an investigator of the Howard Hughes Medical Institute.  

\bibliographystyle{plos}
\bibliography{/Users/bedfordt/Documents/bedford}

\pagebreak

\section*{Tables}

\begin{table}[h]
	\centering
	\caption{\textbf{Rates of mutation and phenotypic change on trunk and side branches and mutational expectation.}}
	\label{mktable}
	\begin{tabular}{ l c c c c } 
	\hline
		 								& Baseline 	& Side branch 	& Trunk		& Trunk / side branch \\
	\hline				
	Mutation size (AG units)			& 0.60		& 0.79			& 1.58		& 1.99$\times$ \\
	Mutation rate (mut per year)		& 0.04		& 0.06			& 0.81		& 13.23$\times$ \\	
	Antigenic flux (AG units per year)	& 0.02		& 0.05			& 1.27		& 26.25$\times$ \\		
	\hline
	\end{tabular}
\end{table}

\pagebreak

\section*{Figures}

\vspace*{\fill}
\vspace*{\fill}

\pagebreak

\vspace*{\fill}
\vspace*{\fill}

\pagebreak

\vspace*{\fill}
\vspace*{\fill}

\pagebreak

\vspace*{\fill}
\vspace*{\fill}

\pagebreak

\vspace*{\fill}
\vspace*{\fill}

\setcounter{figure}{0}
\setcounter{table}{0}
\setcounter{page}{1}
\renewcommand{\thefigure}{Figure~S\arabic{figure}}
\renewcommand{\thetable}{Table~S\arabic{table}}
\renewcommand{\thepage}{S\arabic{page}}
  
\section*{Supporting Information}

\vspace*{\fill}
\begin{figure}[H]
	\centering
	\includegraphics{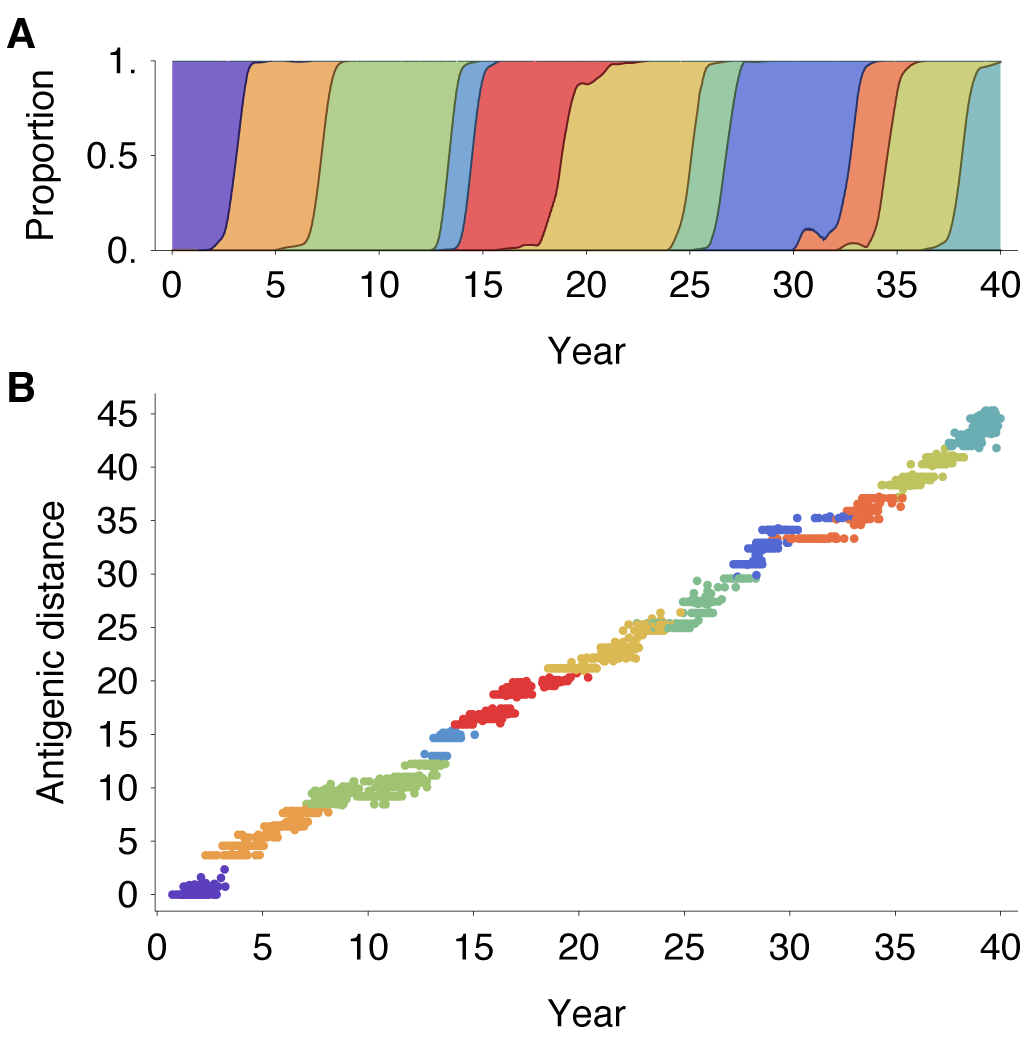}
	\caption{\textbf{Antigenic evolution over the course of the 40-year simulation}. (A) Proportion of virus population comprised of each antigenic cluster through time.  (B) Antigenic distance from initial phenotype ($x=0$, $y=0$) for each of 5943 virus samples relative to time of virus sampling.  Viruses were sampled at a constant rate proportional to prevalence and coloring was determined from the antigenic map in \ref{evol}D.}
	\label{phenotypes}
\end{figure}
\vspace*{\fill}

\pagebreak

\vspace*{\fill}
\begin{figure}[H]
	\centering
	\includegraphics{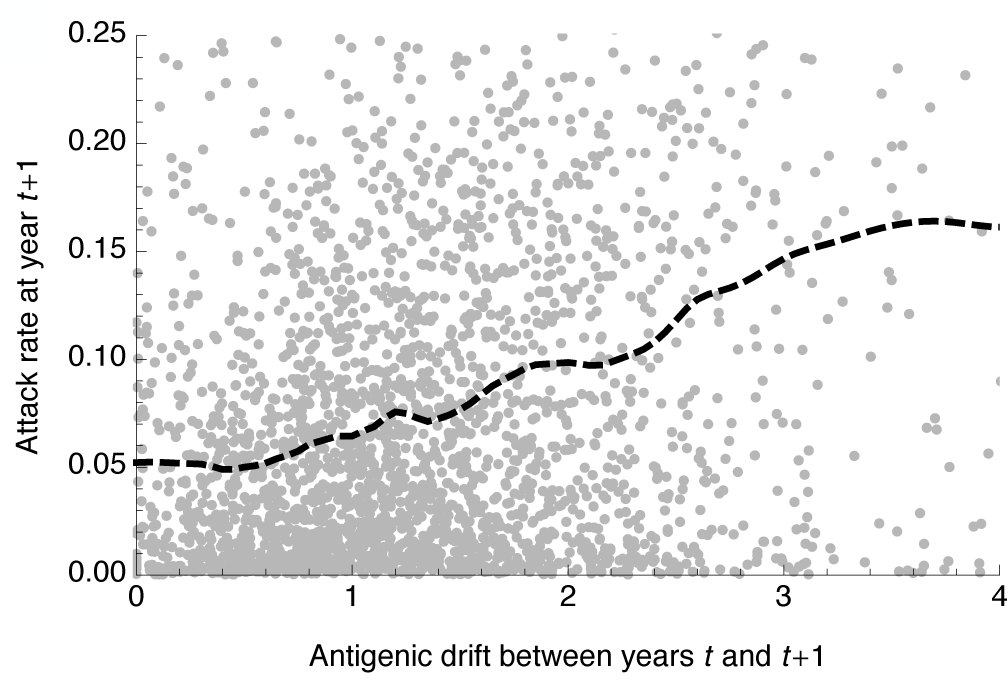}
	\caption{\textbf{Correlation between antigenic drift and attack rate}. Antigenic drift is measured as the distance between the centroid of phenotypes of one year and the centroid of phenotypes of the following year.  Measurements were taken across 80 replicate simulations.  Individual pairs of measurements are shown as gray points and a locally-linear regression (LOESS) is shown as a black dashed line.}
	\label{driftvsinc}
\end{figure}
\vspace*{\fill}

\pagebreak

\vspace*{\fill}
\begin{figure}[H]
	\centering
	\makebox[\textwidth]{
		\includegraphics{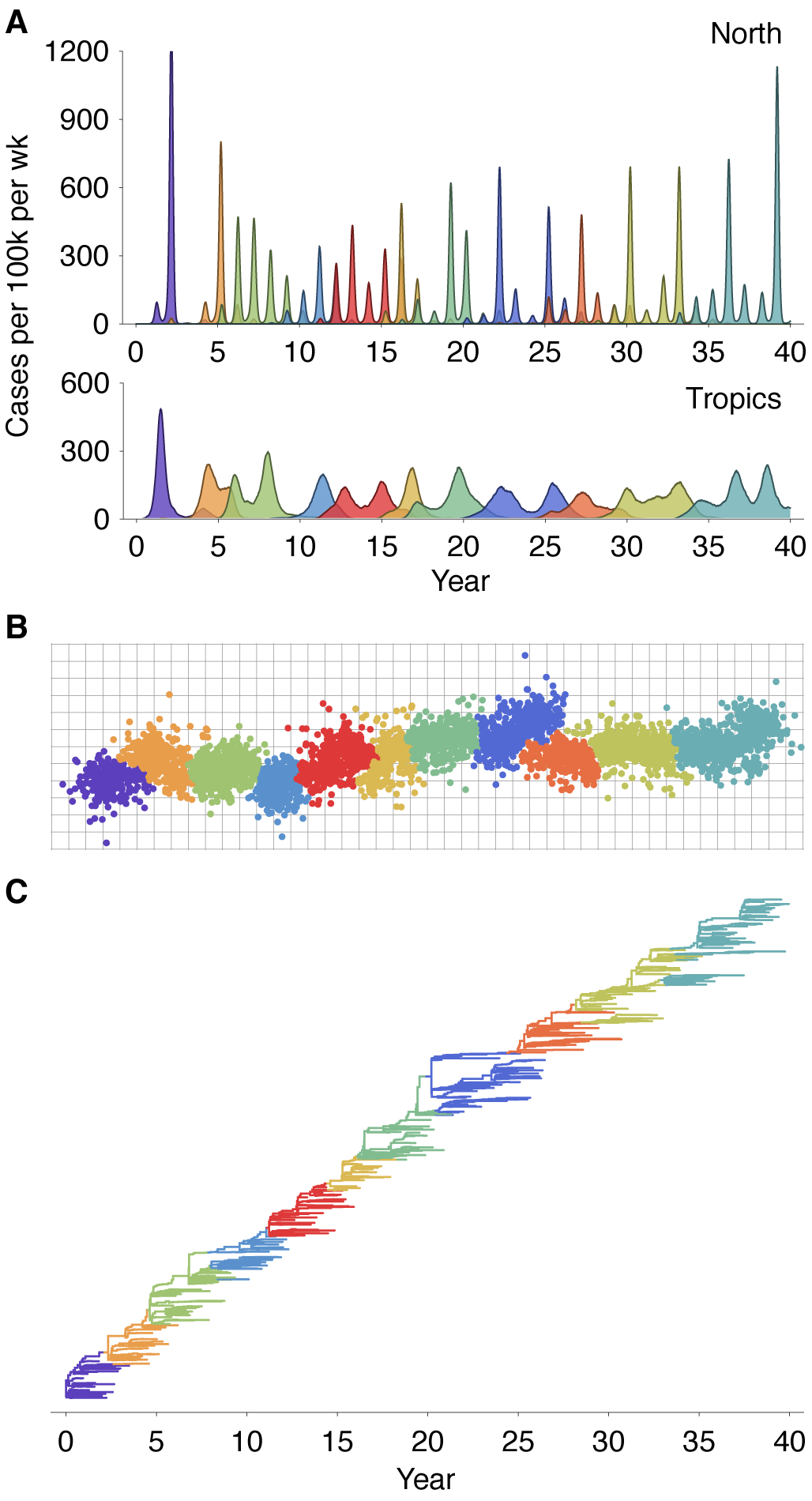}
	}
	\caption{\textbf{Simulation results showing epidemiological, antigenic and genealogical dynamics for `smoother' mutation model}. (A) Weekly timeseries of incidence of viral infection in north and tropics regions. (B) Antigenic map depicting phenotypes of viruses sampled over the course of the simulation.  Grid lines show single units of antigenic distance. (C) Genealogical tree depicting the infection history of samples from the virus population.  Cluster assignments were used to color panels (A), (B) and (C) in a consistent fashion.  Alternative mutational parameters are $\mu = 3 \times 10^{-4}$, mean mutation size of 0.6 units and standard deviation of mutation size of 0.2 units.}
	\label{incmaptree_smooth}
\end{figure}
\vspace*{\fill}

\pagebreak

\vspace*{\fill}
\begin{figure}[H]
	\centering
	\makebox[\textwidth]{
		\includegraphics{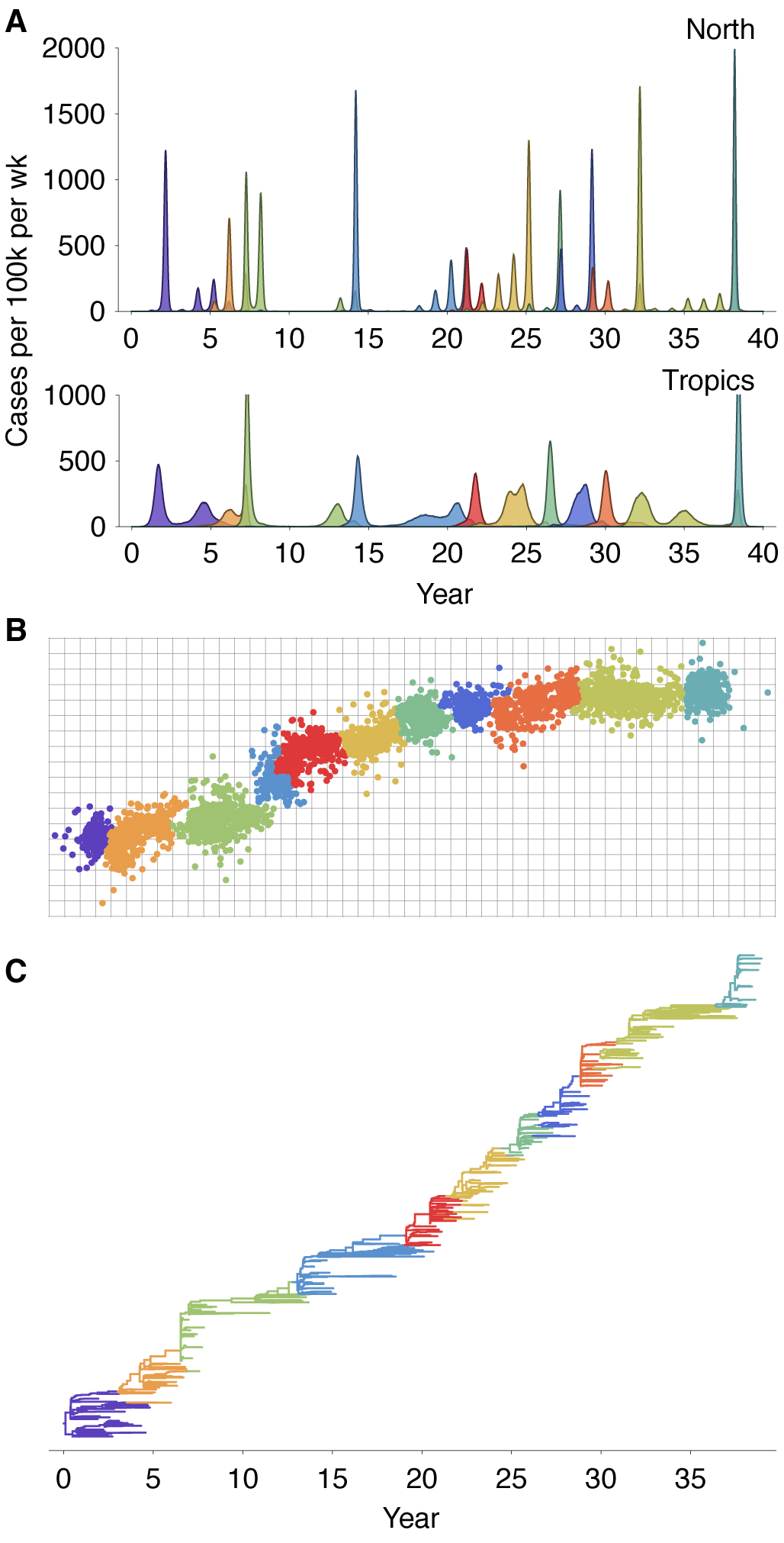}
	}
	\caption{\textbf{Simulation results showing epidemiological, antigenic and genealogical dynamics for `rougher' mutation model}. (A) Weekly timeseries of incidence of viral infection in north and tropics regions. (B) Antigenic map depicting phenotypes of viruses sampled over the course of the simulation.  Grid lines show single units of antigenic distance. (C) Genealogical tree depicting the infection history of samples from the virus population.  Cluster assignments were used to color panels (A), (B) and (C) in a consistent fashion.  Alternative mutational parameters are $\mu = 5 \times 10^{-5}$, mean mutation size of 0.7 units and standard deviation of mutation size of 0.5 units.}
	\label{incmaptree_rough}
\end{figure}
\vspace*{\fill}

\pagebreak

\vspace*{\fill}
\begin{figure}[H]
	\centering
	\makebox[\textwidth]{		
		\includegraphics{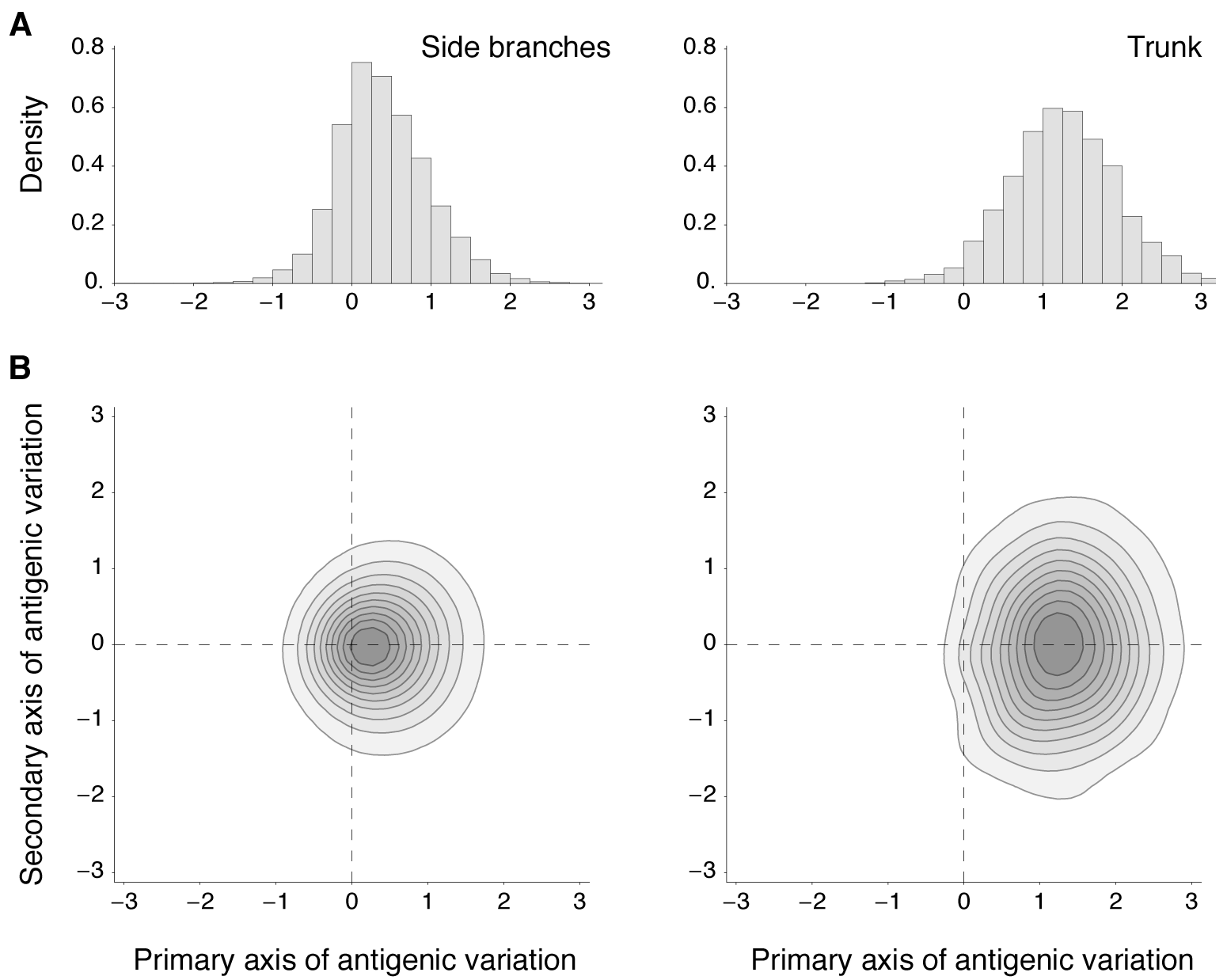}
	}
	\caption{\textbf{Mutation spectrum in two-dimensional antigenic space of side branch mutations and trunk mutations}. (A) Histogram of mutation effects along the axis of primary antigenic variation across 80 replicate simulations.  The left panel shows the distribution of effects of side branch mutations and the right panel shows the distribution of effects of trunk mutations. (B) Smoothed two-dimensional histogram of mutation effects along the primary and secondary axes of antigenic variation across 80 replicate simulations.  Histograms were constructed from 21,405 side branch mutations and 1584 trunk mutations.}
	\label{mutspectrum}
\end{figure}
\vspace*{\fill}

\pagebreak

\vspace*{\fill}
\begin{figure}[H]
	\centering
	\includegraphics{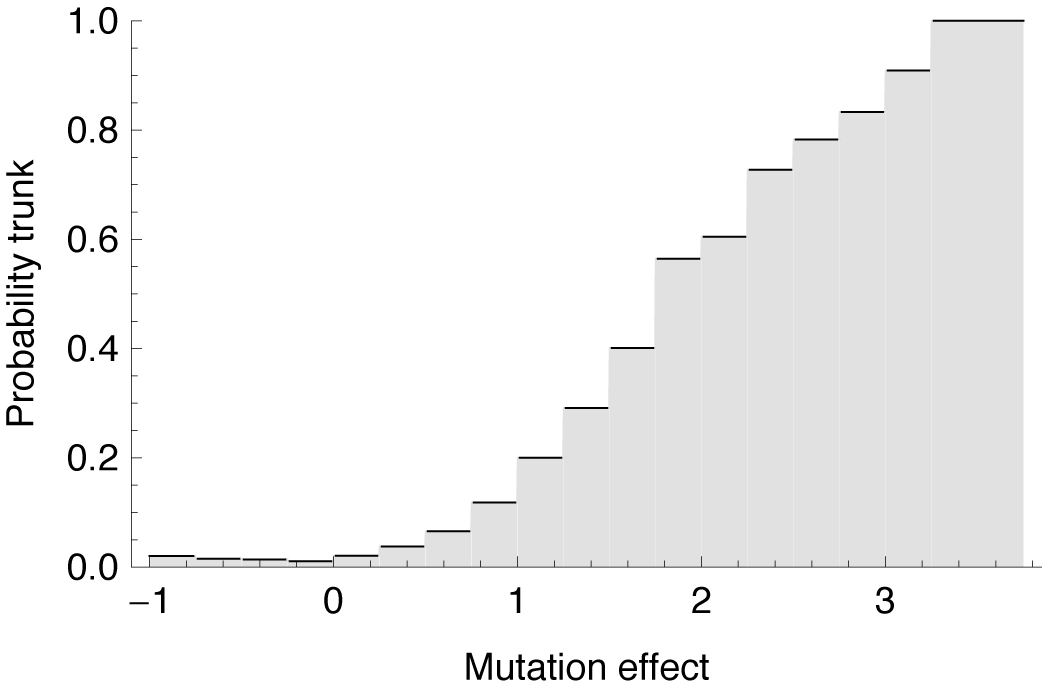}
	\caption{\textbf{Relationship between a mutation's phenotypic effect and its likelihood of being part of the trunk}. The $x$-axis represents the effect of a mutation along the line of primary antigenic variation, and the $y$-axis represents the probability that the mutation is part of the trunk.  Mutations of large effect are increasingly rare, but when they do occur are increasingly likely to be part of the trunk.}
	\label{probtrunk}
\end{figure}
\vspace*{\fill}

\pagebreak

\vspace*{\fill}
\begin{figure}[H]
	\centering
	\includegraphics{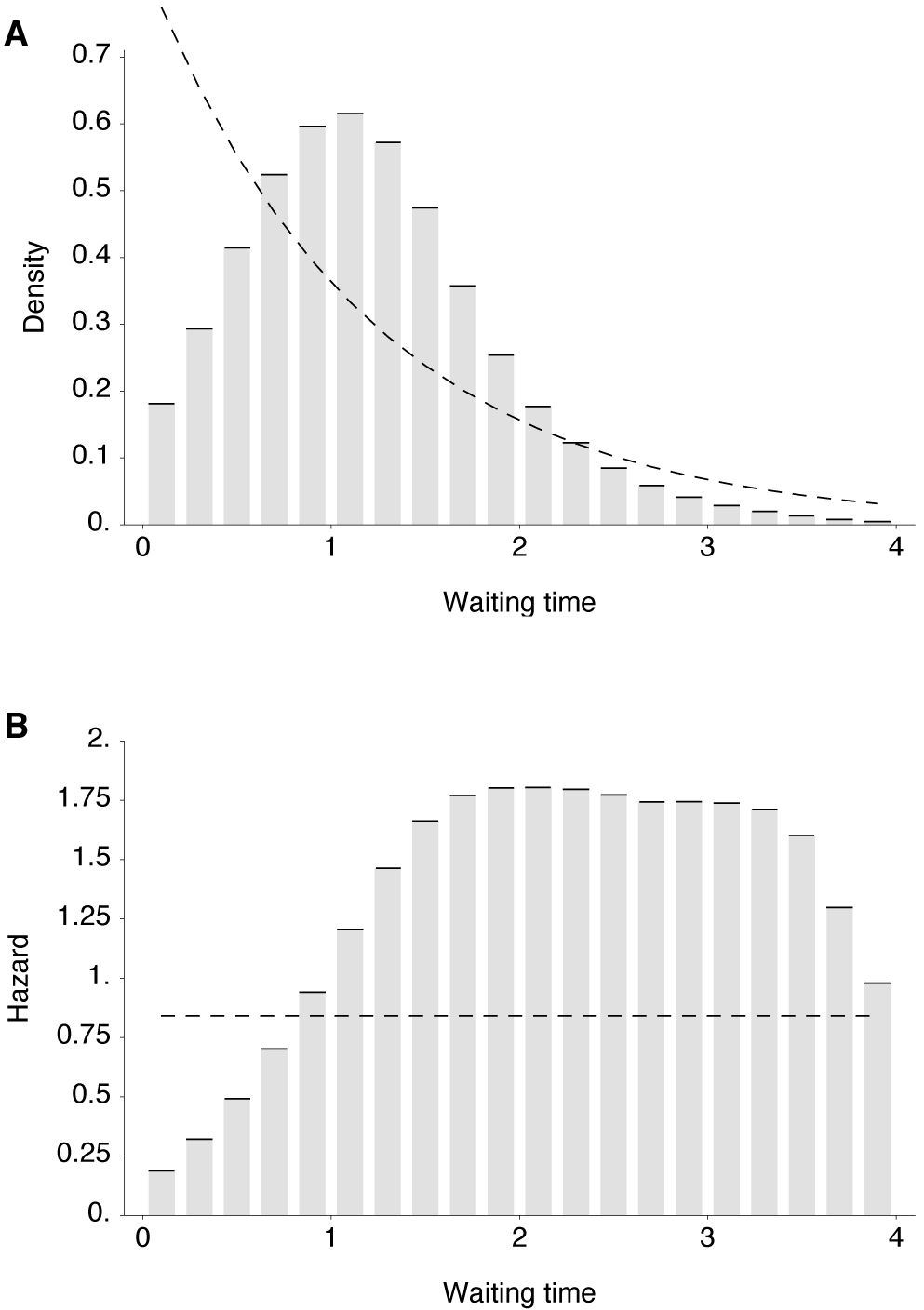}
	\caption{\textbf{Observed vs.\ expected distributions of waiting times between phenotypic mutations along genealogy trunk.} (A) Histogram bins show the observed distribution of waiting times in years across 80 replicate simulations representing 1584 mutations.  The mean of this distribution is 1.76 years.  The dashed line shows the Poisson process expectation of exponentially distributed waiting times.  (B) The density distribution of waiting times is transformed into a hazard function, representing the rate of trunk mutation after a specific waiting time.  The dashed line shows the memoryless hazard function of the Poisson process expectation.}
	\label{waittimes}
\end{figure}
\vspace*{\fill}

\pagebreak

\vspace*{\fill}
\begin{figure}[H]
	\centering
	\makebox[\textwidth]{
		\includegraphics{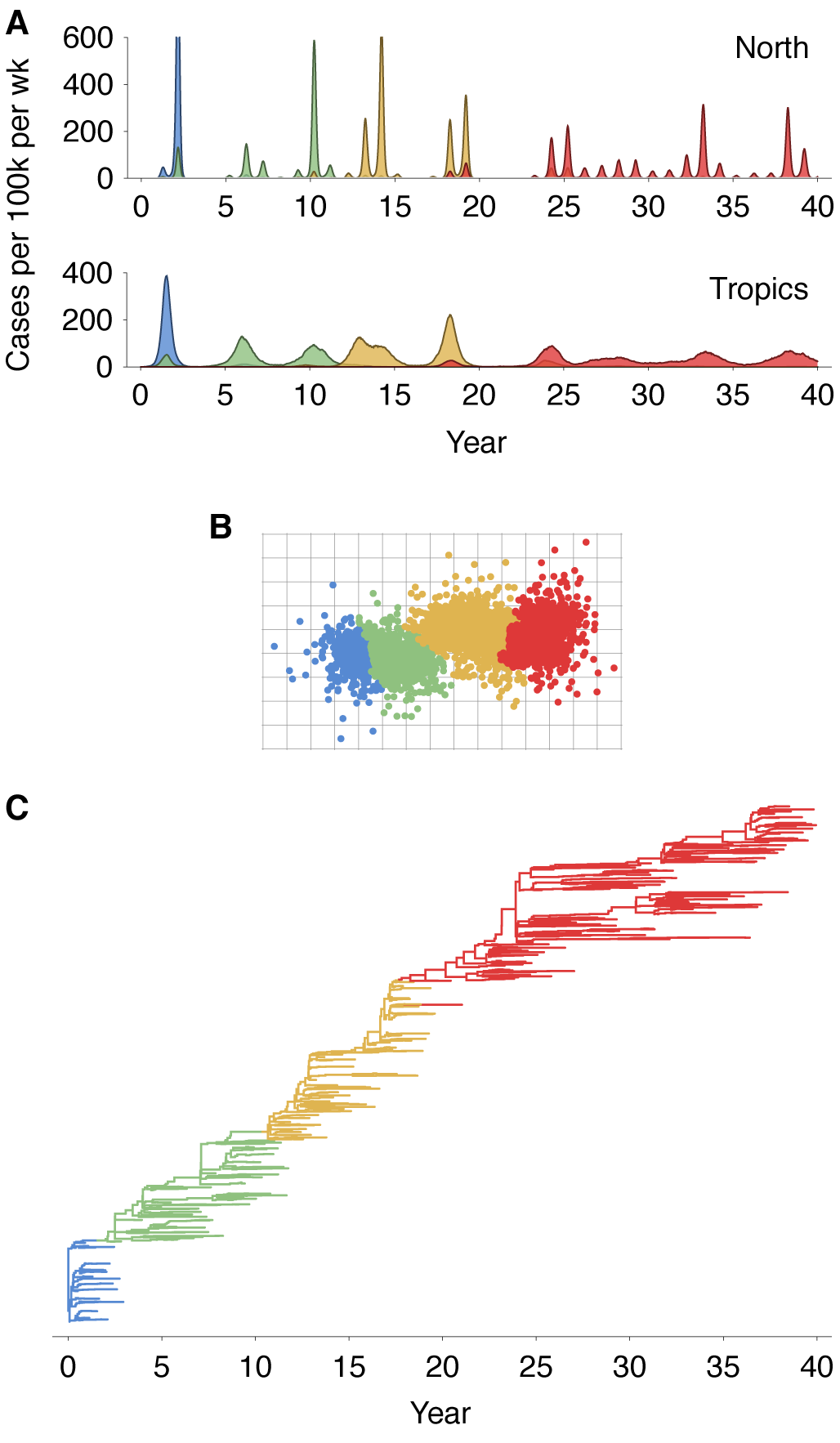}
	}
	\caption{\textbf{Simulation results showing epidemiological, antigenic and genealogical dynamics with weaker mutation}. (A) Weekly timeseries of incidence of viral infection in north and tropics regions. (B) Antigenic map depicting phenotypes of viruses sampled over the course of the simulation.  Grid lines show single units of antigenic distance. (C) Genealogical tree depicting the infection history of samples from the virus population.  Cluster assignments were used to color panels (A), (B) and (C) in a consistent fashion.  Here, $\mu = 5 \times 10^{-5}$, mean mutation size is 0.42 units and standard deviation of mutation size is 0.28 units.}
	\label{h1n1_mut}
\end{figure}
\vspace*{\fill}

\pagebreak

\vspace*{\fill}
\begin{figure}[H]
	\centering
	\makebox[\textwidth]{
		\includegraphics{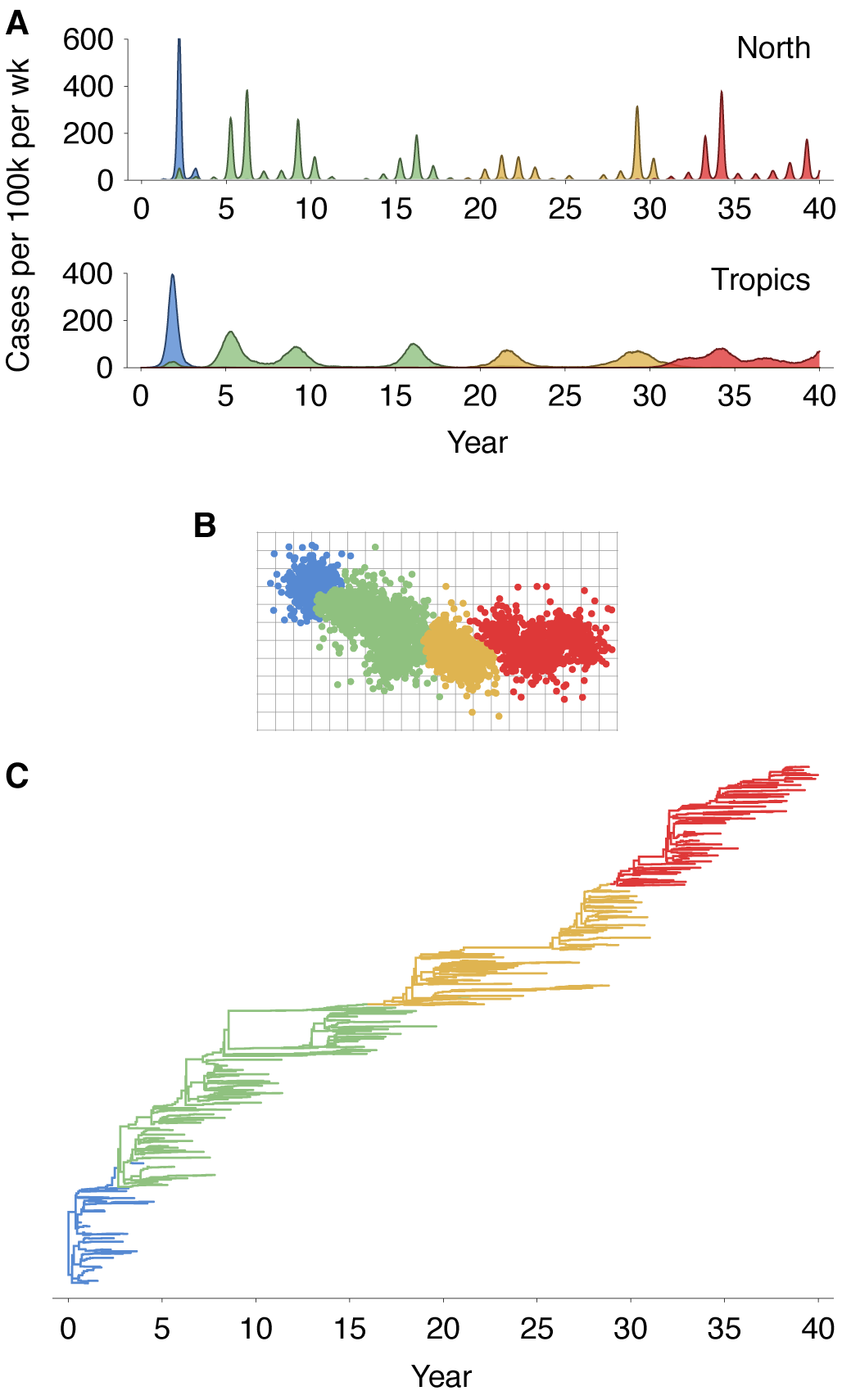}
	}
	\caption{\textbf{Simulation results showing epidemiological, antigenic and genealogical dynamics with lower intrinsic $R_0$}. (A) Weekly timeseries of incidence of viral infection in north and tropics regions. (B) Antigenic map depicting phenotypes of viruses sampled over the course of the simulation.  Grid lines show single units of antigenic distance. (C) Genealogical tree depicting the infection history of samples from the virus population.  Cluster assignments were used to color panels (A), (B) and (C) in a consistent fashion.  Here, $\beta = 0.3$, giving $R_0 = 1.5$.}
	\label{h1n1_r0}
\end{figure}
\vspace*{\fill}

\pagebreak

\vspace*{\fill}
\begin{figure}[H]
	\centering
	\makebox[\textwidth]{	
		\includegraphics[width=0.9\textwidth]{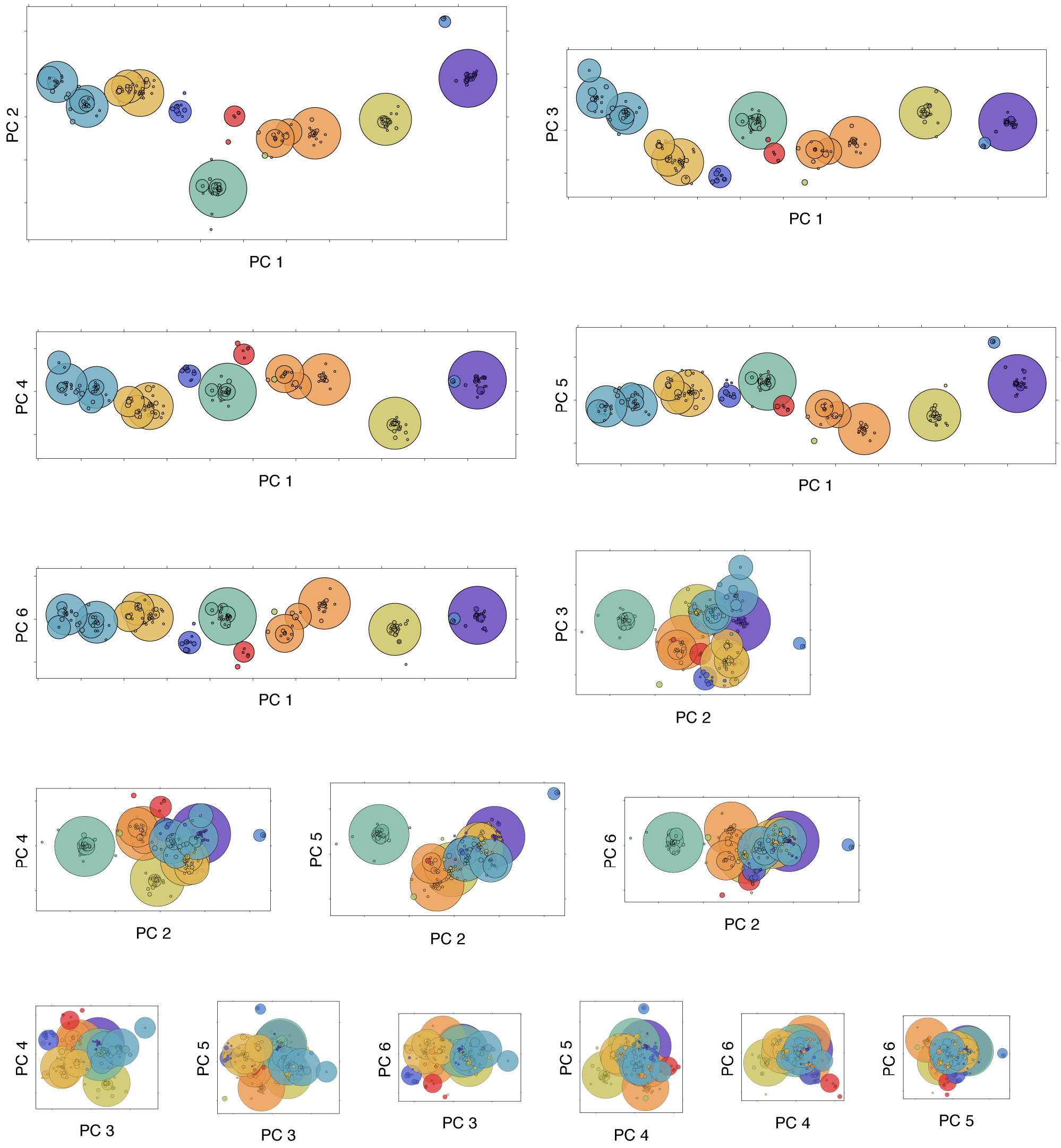}
	}
	\caption{\textbf{Principal components of antigenic variation under a 10-sphere mutation model}. Each panel shows 5991 samples of antigenic phenotype over the course of a 40-year simulation.  Each phenotype is represented as a bubble, with bubble area proportional to the number of samples with this phenotype.  Bubbles are colored based on clustering the 10-dimensional antigenic phenotypes.  The original 10-dimensional space was rotated using principal components analysis to give orthogonal axes in the order of their contribution to antigenic variation.  Each panel shows a two-dimensional slice of the this rotated space.  Principal components 7--10 were left out of the figure for clarity.}
	\label{10dgrid}
\end{figure}
\vspace*{\fill}

\end{document}